\documentclass[english,aip,jcp,reprint,amsmath,amssymb,showpacs,superscriptaddress]{revtex4-1}
\usepackage[T1]{fontenc}
\usepackage{graphicx}
\usepackage{multirow}

\makeatletter
\usepackage{babel}
\makeatother

\begin{document}

\title{Role of three-body interactions in formation of bulk viscosity in liquid argon}

\author{S.~V.~Lishchuk}

\affiliation{Department of Mathematics, University of Leicester, Leicester LE1 7RH, United Kingdom}
\affiliation{School of Food Science and Nutrition, University of Leeds, Leeds LS2 9JT, United Kingdom}


\begin{abstract}

With the aim of locating the origin of discrepancy between experimental and computer simulation
results on bulk viscosity of liquid argon, a molecular dynamic simulation of argon interacting via
ab initio pair potential and triple-dipole three-body potential has been undertaken. Bulk viscosity,
obtained using Green-Kubo formula, is different from the values obtained from modeling argon using
Lennard-Jones potential, the former being closer to the experimental data. The conclusion is made
that many-body inter-atomic interaction plays a significant role in formation of bulk viscosity.

\end{abstract}

\maketitle


\section{Introduction}

Argon above its melting temperature is a typical simple fluid. Consisting of spherical atoms that
interact via short-range repulsion and long-range attraction, and are heavy enough for the quantum
effects to be small, fluid argon and heavier noble gases are an excellent choice of a real system to
be used for testing various approaches in classical theory of fluids.

An inter-particle interaction in argon is commonly represented by a well known 12--6 Lennard-Jones
pair potential \cite{Lennard-Jones:1931-461},
\begin{equation}
\label{eq:Lennard-Jones}
v_{\mathrm{LJ}}(r)=4\epsilon_{\mathrm{LJ}}
\left[\left(\frac{\sigma_{\mathrm{LJ}}}r\right)^{12}
-\left(\frac{\sigma_{\mathrm{LJ}}}r\right)^6\right].
\end{equation}
The two parameters, $\sigma_{\mathrm{LJ}}$ and $\epsilon_{\mathrm{LJ}}$, are usually determined by
fitting thermodynamic properties, derived from the potential (\ref{eq:Lennard-Jones}) by theoretical
or computational methods, to corresponding experimental data.

It is known that Lennard-Jones potential is only an approximation to real interaction in argon.
Several experimental results obtained for argon at large pressures are better explained if a larger
steepness, compared to Lennard-Jones, of argon-argon interaction potential at small inter-atomic
separation distances is taken into account \cite{Bardic:2005-27,Bardic:2006-96}. Accurate
argon--argon interatomic potentials have been calculated by direct {\em ab initio} quantum chemical
calculations \cite{Cybulski:1999-10520,Jaeger:2009-2181,Patkowski:2010-094304} or obtained by
inversion of experimental data \cite{Aziz:1994-5310}. Moreover, many-body dispersion, exchange and
induced polarization contributions to inter-atomic interactions are not small and noticeably
influence thermodynamic properties of argon \cite{Elrod:1994-1975,Jakse:2003-3455}. The most widely
used of these contributions is triple-dipole dispersion interaction, derived by Axilrod and Teller
\cite{Axilrod:1943-299,Axilrod:1951-719} and Muto \cite{Muto:1943-629}, and account of this
contribution in addition to {\em ab initio} pair potential is sufficient to describe thermodynamic
properties of argon with good accuracy
\cite{Barker:1968-134,Anta:1997-2707,Leonhard:2000-1603,Nasrabad:2004-6423,Wang:2006-021202}.

By virtue of Henderson theorem \cite{Henderson:1974-197,Chayes:1984-57}, which states that, for
fluids with only pairwise interactions, and under given conditions of temperature and density, the
pair potential which gives rise to a given radial distribution function $g(r)$ is unique up to a
constant, the thermodynamic properties of the system with many-body interactions can be described by
a model system with an appropriate {\em effective} pair potential. Generally, the effective
potential depends on the thermodynamic state of the system and thermodynamic property to be
described \cite{Casanova:1970-589,Hoef:1999-1520,Louis:2002-9187}. Van der Hoef and Madden
\cite{Hoef:1999-1520} have demonstrated that the account of triple-dipole and
dipole-dipole-quadrupole dispersion interactions moves the effective potential of argon towards
Lennard-Jones form (\ref{eq:Lennard-Jones}). Moreover, the possibility of consistent description of
many thermodynamic properties of argon, using Lennard-Jones potential in a wide domain of
thermodynamic states \cite{Fischer:1984-485,Bembenek:1999-1085,Vrabec:2001-12126}, suggests that the
state dependence of the effective potential is weak.

There is no analogous reason for kinetic properties of a system with many-body interactions to be
equivalent to those of a system with a corresponding effective pair potential. Nevertheless,
experimental data on self-diffusion, shear viscosity and thermal conductivity coefficients of argon
have been shown to be accurately described by Lennard-Jones model with the parameters obtained by
fitting thermodynamic data \cite{Fernandez:2004-175,Fernandez:2004-157}.

Bulk viscosity is a noticeable exception. Bulk viscosity of argon has been measured experimentally
\cite{Naugle:1965-3725,Naugle:1966-4669,Naugle:1966-741,Swyt:1967-1199,Madigosky:1967-4441,Cowan:1972-1881,Baharudin:1975-409,Malbrunot:1983-1523},
and its behavior can be qualitatively described by the results of a molecular dynamics simulation
of a Lennard-Jones system \cite{Meier:2005-014513}. However, when results of simulations with
Lennard-Jones potential are rescaled in an attempt to describe experimental data liquid argon, bulk
viscosity, contrary to other kinetic properties, appears strongly underestimated ({\em e.g.} up to
50\% in Ref.~[\onlinecite{Fernandez:2004-157})].

In view of the above, I propose that the source of this discrepancy may lie in neglect of many-body
interactions. Previous molecular dynamics simulations of systems consisting of 108 particles
interacting via {\em ab initio} pair potential and Axilrod-Teller-Muto (ATM) interaction indicated
that a triple-dipole interaction does not affect the bulk viscosity of liquid xenon near its triple
point \cite{Levesque:1988-918} and dense gaseous krypton \cite{Levesque:1988-3967}. However, the
error in the values of bulk viscosity obtained from molecular dynamics simulation of the systems with
such a small number of particles can be quite large. For example, the values of the reduced bulk
viscosity of the Lennard-Jones systems consisting of 128 and 256 particles at the reduced
temperature $T^*=0.722$ and the reduced density $\rho^*=0.8442$, reported in
Refs~\onlinecite{Heyes:1984-1363,Levesque:1987-143,Hoheisel:1987-7195,Meier:2005-014513}, range from
0.89 to 1.47, with the ratio of the latter to the former of 1.65.

This paper presents the results of more accurate molecular dynamics simulations of a liquid
consisting of 1372 argon atoms with {\em ab initio}+ATM interaction, which demonstrate that bulk
viscosity, determined from Green-Kubo formulae, significantly changes with the account of three-body
interaction, moving results towards experimental data.


\section{Interaction}

Nasrabad {\em et al} \cite{Nasrabad:2004-6423} undertook a Monte Carlo simulation of argon using
combination of {\em ab initio} pair interaction \cite{Cybulski:1999-10520} and ATM triple-dipole
dispersion interaction \cite{Axilrod:1943-299} to test their ability to predict vapor-liquid
equilibrium. Although more accurate {\em ab initio} pair potentials for argon have become available
recently \cite{Jaeger:2009-2181,Patkowski:2010-094304}, and other many-body contributions to
inter-atom interaction can be calculated \cite{Elrod:1994-1975}, we use the same interaction as
Nasrabad {\em et al} because, being able to predict accurately the phase diagram of argon
\cite{Nasrabad:2004-6423}, it is computationally more efficient.

Specifically, the {\em ab initio} pair interaction potential used in the present work is described
by a function \cite{Nasrabad:2004-6423}
\begin{equation}
\label{eq:ab-initio}
u_2(r)=Ae^{-\alpha r+\beta r^2}+\sum_{n=3}^5f_{2n}(r,b)\frac{C_{2n}}{r^{2n}},
\end{equation}
where
\begin{equation}
f_{2n}(r,b)=1-e^{-br}\sum_{k=0}^{2n}\frac{(br)^k}{k!},
\end{equation}
and numerical values of the parameters $A$, $\alpha$, $\beta$, $b$, and $C_{2n}$ are given in
Ref.~[\onlinecite{Nasrabad:2004-6423}]. The ATM triple-dipole interaction has form
\cite{Axilrod:1943-299}
\begin{equation}
\label{eq:triple-dipole}
u_3(r_{12},r_{23},r_{31})=\nu\frac{1+3\cos\alpha\cos\beta\cos\gamma}{r_{12}^3r_{23}^3r_{31}^3},
\end{equation}
where the $r_{ik}$ are the lengths of the sides, $\alpha$, $\beta$, and $\gamma$ are the angles of
the triangle formed by three argon atoms, and $\nu=7.32\cdot10^{-108}$\,J$\cdot$m${}^9$ for argon
\cite{Barker:1968-134,Anta:1997-2707}.

For simulations of argon  using Lennard-Jones potential (\ref{eq:Lennard-Jones}) the values
$\sigma_{\mathrm{LJ}}=3.3952$\,\AA{} and $\epsilon_{\mathrm{LJ}}=116.79$\,K are used
\cite{Vrabec:2001-12126}.


\section{Simulation}

Meier {\em et al} \cite{Meier:2005-014513} undertook a systematic study of the influence of the
number of particles and the cutoff radius for pair interaction on the bulk viscosity of
Lennard-Jones system. In view of their results, simulations were performed in a cubic box containing
$N=1372$ particles, and the cutoff radius for pair interactions was set to $5\sigma_{\mathrm{LJ}}$.
Three-body interactions were cut off when the distance between any pair of the atoms in the triplet
exceeded one quarter of the simulation box length (around $3\sigma_{\mathrm{LJ}}$ for the densities
studied in this work). Usual periodic boundary conditions and minimum image convention were applied.
The simulations were started with the particles in a face-centered-cubic lattice, with randomly
assigned velocities. Forces arising from three-body interactions were calculated using formulas
given by Allen and Tildesley \cite{Allen:CSL}, and an expression for forces due to {\em ab initio}
pair interaction was obtained by applying gradient operator to Eq.~(\ref{eq:ab-initio}). Newton's
equations of motion were solved using velocity-Verlet algorithm with the time step
$\Delta t\cdot\sqrt{\epsilon_{\mathrm{LJ}}/m}/\sigma_{\mathrm{LJ}}=0.003$.

The runs were made at the experimental densities at various temperatures along the 40~atm isochore,
taken from Ref.~[\onlinecite{Cowan:1972-1881}]. Every simulation was initiated in the NVT ensemble
and run for at least $2{\cdot}10^5$ time steps to attain thermodynamic equilibrium.
After equilibration the thermostat was turned off and the NVE ensemble was invoked to calculate bulk
and shear viscosities. The length of the production period was $4{\cdot}10^6$ time steps for the
system interacting via Lennard-Jones potential, and between $10^6$ and $3{\cdot}10^6$ time steps for
the system with {\em ab initio}\,+\,ATM interaction, depending on the state point.

Bulk viscosity, $\zeta$, and shear viscosity, $\eta$, were calculated using Green-Kubo formulas
\cite{Green:1954-398}:
\begin{equation}
\label{eq:zeta}
\zeta=\frac V{k_BT}\int_0^\infty\left<\delta p(t)\delta p(t_0)\right>dt,
\end{equation}
\begin{equation}
\label{eq:eta}
\eta=\frac V{k_BT}\int_0^\infty\left<\sigma_{\alpha\beta}(t)\sigma_{\alpha\beta}(t_0)\right>dt,
\end{equation}
where $V$ is volume, $k_B$ is Boltzmann constant, $T$ is temperature, $t$ is time,
$\delta p=p-\left<p\right>$ is the deviation of the instantaneous pressure $p$ from its average
value $\left<p\right>$, $\sigma_{\alpha\beta}$ is an off-diagonal element of the stress tensor, the
angular brackets denote equilibrium ensemble averages over short trajectory sections of the
phase-space trajectory of the system with multiple (every time step) time origins $t_0$. The stress
tensor was calculated using formulae given by Lee and Cummings \cite{Lee:1994-6206}. The integration
in Eqs~(\ref{eq:zeta}) and (\ref{eq:eta}) was carried out up to $\tau_L=L/c$, where $L$ is
simulation box length and $c$ is sound velocity taken from Ref.~[\onlinecite{Cowan:1972-1881}].
Depending on the state point, the value of $\tau_L$ was between 4.80 and 11.25~ps. The statistical
error in time correlation functions was estimated using formula given by Frenkel and Smit
\cite{Frenkel:UMS},
\begin{equation}
\label{eq:error}
\sigma\left(\left<X(t)X(0)\right>\right)
\approx\sqrt{\frac{2\tau_X}{t_{\mathrm{run}}}}\left<X^2(0)\right>,
\end{equation}
where $t_{\mathrm{run}}$ is the length of the simulation, and the correlation time $\tau_X$ was
approximated as the time during which time correlation function decays $e\approx2.718$ times.


\section{Results}

\begin{figure}[tb]
\begin{center}
\includegraphics[width=\columnwidth,keepaspectratio]{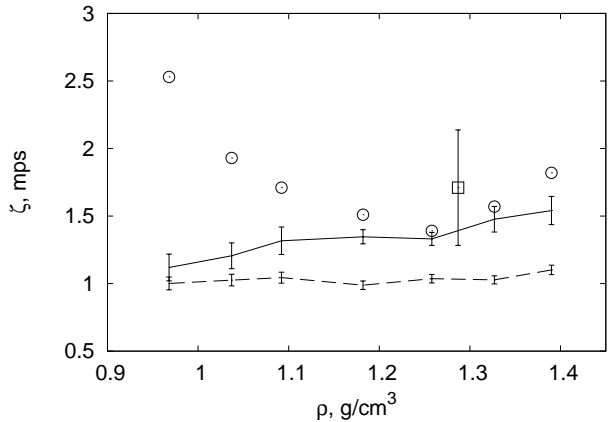}
\end{center}
\caption{
Bulk viscosity of liquid argon at $T=(90{-}140)$\,K. Error bars connected with solid and dashed
lines correspond to the simulation results with {\em ab initio}\,+\,ATM and Lennard-Jones
interaction, respectively. Experimental points are taken from Refs~[\onlinecite{Cowan:1972-1881}]
(circles, pressure 40~atm) and [\onlinecite{Naugle:1966-4669}] (square with error bar, pressure
40~kg/cm${}^2$).
}
\label{fig:zeta}
\end{figure}

\begin{figure}[tb]
\begin{center}
\includegraphics[width=\columnwidth,keepaspectratio]{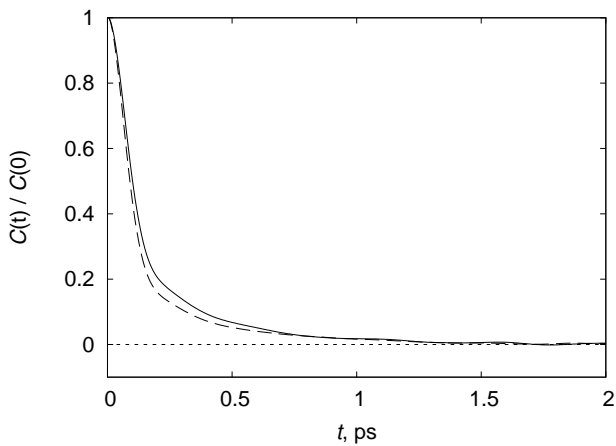}
\end{center}
\caption{
Time-correlation functions $C(t)$ used for calculation of bulk viscosity at density
1.258\,g/cm${}^3$. Solid and dashed lines correspond to the simulation results with
{\em ab initio}\,+\,ATM and Lennard-Jones  interaction, respectively.
}
\label{fig:tcf}
\end{figure}

Fig.~\ref{fig:zeta} and Table~\ref{table:results} present simulation results for the bulk viscosity
obtained using {\em ab initio}\,+\,ATM (Eqs~(\ref{eq:ab-initio}) and (\ref{eq:triple-dipole})) and
Lennard-Jones (Eq.~(\ref{eq:Lennard-Jones})) interaction, respectively. Bulk viscosity, determined
from Green-Kubo formulas, changes with the account of three-body interaction, moving towards
experimental data. However, this change is not sufficient to obtain numerical agreement with
experiment, especially at lower densities. Typical behavior of time correlation functions
$C(t)=\left<\delta p(t)\delta p(0)\right>$ is shown in Fig.~\ref{fig:tcf}.

\begin{table*}[tb]
\begin{center}
\begin{tabular}{|c|c|c|c|c|c|c|c|c|}
\hline
\multirow{2}{*}{~$T$, K~} &
\multirow{2}{*}{$\rho$, g/cm${}^3$} &
\multicolumn{3}{|c|}{Bulk viscosity $\zeta$, mps} &
\multicolumn{4}{|c|}{Shear viscosity $\eta$, mps} \\
\cline{3-9}
& &
LJ & AI+ATM & Ref.~[\onlinecite{Cowan:1972-1881}] &
LJ & AI+ATM & Ref.~[\onlinecite{Cowan:1972-1881}] & Ref.~[\onlinecite{NIST-old}] \\
\hline
~90 & 1.390 & $1.10\pm0.04$ & $1.54\pm0.10$ & 1.82 & $2.31\pm0.04$ & $2.44\pm0.07$ & ~2.33~ & 2.57 \\
\hline
100 & 1.327 & $1.03\pm0.03$ & $1.48\pm0.09$ & 1.57 & $1.78\pm0.03$ & $1.87\pm0.06$ & ~1.86~ & 1.92 \\
\hline
110 & 1.258 & $1.04\pm0.02$ & $1.33\pm0.05$ & 1.39 & $1.38\pm0.02$ & $1.39\pm0.03$ & ~1.51~ & 1.48 \\
\hline
120 & 1.182 & $0.99\pm0.03$ & $1.35\pm0.05$ & 1.51 & $1.09\pm0.02$ & $1.12\pm0.02$ & ~1.19~ & 1.15 \\
\hline
130 & 1.092 & $1.04\pm0.04$ & $1.32\pm0.10$ & 1.71 & $0.86\pm0.02$ & $0.87\pm0.03$ & ~0.88~ & 0.89 \\
\hline
135 & 1.037 & $1.03\pm0.04$ & $1.21\pm0.10$ & 1.93 & $0.73\pm0.02$ & $0.70\pm0.03$ & ~0.760 & 0.77 \\
\hline
140 & 0.968 & $1.00\pm0.05$ & $1.12\pm0.10$ & 2.53 & $0.65\pm0.02$ & $0.65\pm0.03$ & ~0.642 & 0.65 \\
\hline
\end{tabular}
\end{center}
\caption{
Bulk and shear viscosities of argon obrained from molecular dynamics simulations using Lennard-Jones
(LJ) and {\em ab initio pair}~+~Axilrod-Teller-Muto three-body (AI+ATM) interaction, and
corresponding experimental data \cite{Cowan:1972-1881,NIST-old}. Error in the simulation data is
calculated using Eq.~(\ref{eq:error}).
}
\label{table:results}
\end{table*}

\begin{figure}[tb]
\begin{center}
\includegraphics[width=\columnwidth,keepaspectratio]{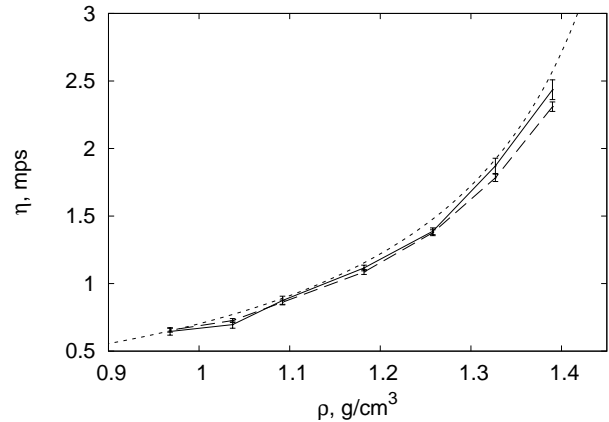}
\end{center}
\caption{
Shear viscosity of liquid argon at $T=(90{-}140)$\,K. Error bars connected with solid and dashed
lines correspond to the simulation results with {\em ab initio}\,+\,ATM and Lennard-Jones
interaction, respectively. Dotted line corresponds to the interpolation data for pressure 40~atm
taken from Ref.~[\onlinecite{NIST-old}].
}
\label{fig:eta}
\end{figure}

Fernandez {\em et al} \cite{Fernandez:2004-157} demonstrated that, contrary to bulk viscosity, the
values of shear viscosity of argon obtained from molecular dynamics simulation of a Lennard-Jones
system agree with experimental data. Lee and Cummings \cite{Lee:1994-6206} and Marcelli {\em et al}
\cite{Marcelli:2001-021204} found that the influence of triple-dipole interaction on shear viscosity
of argon is small. The results of the present simulation, shown in Fig.~\ref{fig:eta} and
Table~\ref{table:results}, agree with these findings.

\section{Conclusion}

The message of this paper is that many-body interactions play a more substantial role in determining
the value of the bulk viscosity than other transport coefficients. The present results from the
molecular dynamic simulation of liquid argon demonstrate that even account of a single many-body
contribution, ATM triple-dipole interaction, shifts the values of the bulk viscosity of argon
towards experimental data. Larger sensitivity of the bulk viscosity to many-body interaction,
compared to other transport coefficients, can be intuitively explained in the case of gaseous state.
Bulk viscosity of a non-relativistic monoatomic gas calculated from the Boltzmann equation, which
takes into account only pair collisions of atoms, appears to be zero, in contrast to heat
conductivity and shear viscosity which have non-zero values in the same approximation
\cite{Lifshitz:PK}. A non-zero value of bulk viscosity appears in the approximations corresponding
to higher-order terms in the virial expansion \cite{Choh:KTPDG,Baimbetov:1984-1781}, which
correspond to the explicit account of at least three-atom collisions which, in turn, are sensitive
to three-body inter-atomic interaction.


\begin{acknowledgments}
I thank
Prof.\ Jadran Vrabec, Dr Andrey Brukhno, and Dr Ian Halliday
for stimulating discussions.
\end{acknowledgments}



\begin{thebibliography}{50}
\expandafter\ifx\csname natexlab\endcsname\relax\def\natexlab#1{#1}\fi
\expandafter\ifx\csname bibnamefont\endcsname\relax
  \def\bibnamefont#1{#1}\fi
\expandafter\ifx\csname bibfnamefont\endcsname\relax
  \def\bibfnamefont#1{#1}\fi
\expandafter\ifx\csname citenamefont\endcsname\relax
  \def\citenamefont#1{#1}\fi
\expandafter\ifx\csname url\endcsname\relax
  \def\url#1{\texttt{#1}}\fi
\expandafter\ifx\csname urlprefix\endcsname\relax\def\urlprefix{URL }\fi
\providecommand{\bibinfo}[2]{#2}
\providecommand{\eprint}[2][]{\url{#2}}

\bibitem[{\citenamefont{{Lennard-Jones}}(1931)}]{Lennard-Jones:1931-461}
\bibinfo{author}{\bibfnamefont{J.~E.} \bibnamefont{{Lennard-Jones}}},
  \bibinfo{journal}{Proc.\ Phys.\ Soc.} \textbf{\bibinfo{volume}{43}},
  \bibinfo{pages}{461} (\bibinfo{year}{1931}).

\bibitem[{\citenamefont{Bardic et~al.}(2005)\citenamefont{Bardic, Malomuzh, and
  Sysoev}}]{Bardic:2005-27}
\bibinfo{author}{\bibfnamefont{V.~Y.} \bibnamefont{Bardic}},
  \bibinfo{author}{\bibfnamefont{N.~P.} \bibnamefont{Malomuzh}},
  \bibnamefont{and} \bibinfo{author}{\bibfnamefont{V.~M.}
  \bibnamefont{Sysoev}}, \bibinfo{journal}{J.\ Mol.\ Liquids}
  \textbf{\bibinfo{volume}{120}}, \bibinfo{pages}{27} (\bibinfo{year}{2005}).

\bibitem[{\citenamefont{Bardic et~al.}(2006)\citenamefont{Bardic, Malomuzh,
  Shakun, and Sysoev}}]{Bardic:2006-96}
\bibinfo{author}{\bibfnamefont{V.~Y.} \bibnamefont{Bardic}},
  \bibinfo{author}{\bibfnamefont{N.~P.} \bibnamefont{Malomuzh}},
  \bibinfo{author}{\bibfnamefont{K.~S.} \bibnamefont{Shakun}},
  \bibnamefont{and} \bibinfo{author}{\bibfnamefont{V.~M.}
  \bibnamefont{Sysoev}}, \bibinfo{journal}{J.\ Mol.\ Liquids}
  \textbf{\bibinfo{volume}{127}}, \bibinfo{pages}{96} (\bibinfo{year}{2006}).

\bibitem[{\citenamefont{Cybulski and
  Toczy\l{}owski}(1999)}]{Cybulski:1999-10520}
\bibinfo{author}{\bibfnamefont{S.~M.} \bibnamefont{Cybulski}} \bibnamefont{and}
  \bibinfo{author}{\bibfnamefont{R.~R.} \bibnamefont{Toczy\l{}owski}},
  \bibinfo{journal}{J.\ Chem.\ Phys.} \textbf{\bibinfo{volume}{111}},
  \bibinfo{pages}{10520} (\bibinfo{year}{1999}).

\bibitem[{\citenamefont{J\"ager et~al.}(2009)\citenamefont{J\"ager, Hellmann,
  Bich, and Vogel}}]{Jaeger:2009-2181}
\bibinfo{author}{\bibfnamefont{B.}~\bibnamefont{J\"ager}},
  \bibinfo{author}{\bibfnamefont{R.}~\bibnamefont{Hellmann}},
  \bibinfo{author}{\bibfnamefont{E.}~\bibnamefont{Bich}}, \bibnamefont{and}
  \bibinfo{author}{\bibfnamefont{E.}~\bibnamefont{Vogel}},
  \bibinfo{journal}{Mol.\ Phys.} \textbf{\bibinfo{volume}{107}},
  \bibinfo{pages}{2181} (\bibinfo{year}{2009}).

\bibitem[{\citenamefont{Patkowski and Szalewicz}(2010)}]{Patkowski:2010-094304}
\bibinfo{author}{\bibfnamefont{K.}~\bibnamefont{Patkowski}} \bibnamefont{and}
  \bibinfo{author}{\bibfnamefont{K.}~\bibnamefont{Szalewicz}},
  \bibinfo{journal}{J.\ Chem.\ Phys.} \textbf{\bibinfo{volume}{133}},
  \bibinfo{pages}{094304} (\bibinfo{year}{2010}).

\bibitem[{\citenamefont{Aziz et~al.}(1994)\citenamefont{Aziz, Slaman, and
  Janzen}}]{Aziz:1994-5310}
\bibinfo{author}{\bibfnamefont{R.~A.} \bibnamefont{Aziz}},
  \bibinfo{author}{\bibfnamefont{M.~J.} \bibnamefont{Slaman}},
  \bibnamefont{and} \bibinfo{author}{\bibfnamefont{A.~R.}
  \bibnamefont{Janzen}}, \bibinfo{journal}{Phys.\ Rev.\ E}
  \textbf{\bibinfo{volume}{49}}, \bibinfo{pages}{5310} (\bibinfo{year}{1994}).

\bibitem[{\citenamefont{Elrod and Saykally}(1994)}]{Elrod:1994-1975}
\bibinfo{author}{\bibfnamefont{M.~J.} \bibnamefont{Elrod}} \bibnamefont{and}
  \bibinfo{author}{\bibfnamefont{R.~J.} \bibnamefont{Saykally}},
  \bibinfo{journal}{Chem.\ Rev.} \textbf{\bibinfo{volume}{94}},
  \bibinfo{pages}{1975} (\bibinfo{year}{1994}).

\bibitem[{\citenamefont{Jakse and Bretonnet}(2003)}]{Jakse:2003-3455}
\bibinfo{author}{\bibfnamefont{N.}~\bibnamefont{Jakse}} \bibnamefont{and}
  \bibinfo{author}{\bibfnamefont{J.-L.} \bibnamefont{Bretonnet}},
  \bibinfo{journal}{J.\ Phys.\ Cond.\ Matter} \textbf{\bibinfo{volume}{15}},
  \bibinfo{pages}{S3455} (\bibinfo{year}{2003}).

\bibitem[{\citenamefont{Axilrod and Teller}(1943)}]{Axilrod:1943-299}
\bibinfo{author}{\bibfnamefont{B.~M.} \bibnamefont{Axilrod}} \bibnamefont{and}
  \bibinfo{author}{\bibfnamefont{E.}~\bibnamefont{Teller}},
  \bibinfo{journal}{J.\ Chem.\ Phys.} \textbf{\bibinfo{volume}{11}},
  \bibinfo{pages}{299} (\bibinfo{year}{1943}).

\bibitem[{\citenamefont{Axilrod}(1951)}]{Axilrod:1951-719}
\bibinfo{author}{\bibfnamefont{B.~M.} \bibnamefont{Axilrod}},
  \bibinfo{journal}{J.\ Chem.\ Phys.} \textbf{\bibinfo{volume}{19}},
  \bibinfo{pages}{719} (\bibinfo{year}{1951}).

\bibitem[{\citenamefont{Muto}(1943)}]{Muto:1943-629}
\bibinfo{author}{\bibfnamefont{Y.}~\bibnamefont{Muto}},
  \bibinfo{journal}{Proc.\ Phys.\ Math.\ Soc.\ Japan}
  \textbf{\bibinfo{volume}{17}}, \bibinfo{pages}{629} (\bibinfo{year}{1943}).

\bibitem[{\citenamefont{Barker et~al.}(1968)\citenamefont{Barker, Henderson,
  and Smith}}]{Barker:1968-134}
\bibinfo{author}{\bibfnamefont{J.~A.} \bibnamefont{Barker}},
  \bibinfo{author}{\bibfnamefont{D.}~\bibnamefont{Henderson}},
  \bibnamefont{and} \bibinfo{author}{\bibfnamefont{W.~R.} \bibnamefont{Smith}},
  \bibinfo{journal}{Phys.\ Rev.\ Lett.} \textbf{\bibinfo{volume}{21}},
  \bibinfo{pages}{134} (\bibinfo{year}{1968}).

\bibitem[{\citenamefont{Anta et~al.}(1997)\citenamefont{Anta, Lomba, and
  Lombardero}}]{Anta:1997-2707}
\bibinfo{author}{\bibfnamefont{J.~A.} \bibnamefont{Anta}},
  \bibinfo{author}{\bibfnamefont{E.}~\bibnamefont{Lomba}}, \bibnamefont{and}
  \bibinfo{author}{\bibfnamefont{M.}~\bibnamefont{Lombardero}},
  \bibinfo{journal}{Phys.\ Rev.\ E} \textbf{\bibinfo{volume}{55}},
  \bibinfo{pages}{2707} (\bibinfo{year}{1997}).

\bibitem[{\citenamefont{Leonhard and Deiters}(2000)}]{Leonhard:2000-1603}
\bibinfo{author}{\bibfnamefont{K.}~\bibnamefont{Leonhard}} \bibnamefont{and}
  \bibinfo{author}{\bibfnamefont{U.~K.} \bibnamefont{Deiters}},
  \bibinfo{journal}{Mol.\ Phys.} \textbf{\bibinfo{volume}{98}},
  \bibinfo{pages}{1603} (\bibinfo{year}{2000}).

\bibitem[{\citenamefont{Nasrabad et~al.}(2004)\citenamefont{Nasrabad, Laghaei,
  and Deiters}}]{Nasrabad:2004-6423}
\bibinfo{author}{\bibfnamefont{A.~E.} \bibnamefont{Nasrabad}},
  \bibinfo{author}{\bibfnamefont{R.}~\bibnamefont{Laghaei}}, \bibnamefont{and}
  \bibinfo{author}{\bibfnamefont{U.~K.} \bibnamefont{Deiters}},
  \bibinfo{journal}{J.\ Chem.\ Phys.} \textbf{\bibinfo{volume}{121}},
  \bibinfo{pages}{6423} (\bibinfo{year}{2004}).

\bibitem[{\citenamefont{Wang and Sadus}(2006)}]{Wang:2006-021202}
\bibinfo{author}{\bibfnamefont{L.}~\bibnamefont{Wang}} \bibnamefont{and}
  \bibinfo{author}{\bibfnamefont{R.~J.} \bibnamefont{Sadus}},
  \bibinfo{journal}{Phys.\ Rev.\ E} \textbf{\bibinfo{volume}{74}},
  \bibinfo{pages}{021202} (\bibinfo{year}{2006}).

\bibitem[{\citenamefont{Henderson}(1974)}]{Henderson:1974-197}
\bibinfo{author}{\bibfnamefont{R.~L.} \bibnamefont{Henderson}},
  \bibinfo{journal}{Phys.\ Lett.\ A} \textbf{\bibinfo{volume}{49}},
  \bibinfo{pages}{197} (\bibinfo{year}{1974}).

\bibitem[{\citenamefont{Chayes et~al.}(1984)\citenamefont{Chayes, Chayes, and
  Lieb}}]{Chayes:1984-57}
\bibinfo{author}{\bibfnamefont{J.~T.} \bibnamefont{Chayes}},
  \bibinfo{author}{\bibfnamefont{L.}~\bibnamefont{Chayes}}, \bibnamefont{and}
  \bibinfo{author}{\bibfnamefont{E.~H.} \bibnamefont{Lieb}},
  \bibinfo{journal}{Commun.\ Math.\ Phys.} \textbf{\bibinfo{volume}{93}},
  \bibinfo{pages}{57} (\bibinfo{year}{1984}).

\bibitem[{\citenamefont{Casanova et~al.}(1970)\citenamefont{Casanova, Dulla,
  Jonah, Rowlinson, and Saville}}]{Casanova:1970-589}
\bibinfo{author}{\bibfnamefont{G.}~\bibnamefont{Casanova}},
  \bibinfo{author}{\bibfnamefont{R.~J.} \bibnamefont{Dulla}},
  \bibinfo{author}{\bibfnamefont{D.~A.} \bibnamefont{Jonah}},
  \bibinfo{author}{\bibfnamefont{J.~S.} \bibnamefont{Rowlinson}},
  \bibnamefont{and} \bibinfo{author}{\bibfnamefont{G.}~\bibnamefont{Saville}},
  \bibinfo{journal}{Mol.\ Phys.} \textbf{\bibinfo{volume}{18}},
  \bibinfo{pages}{589} (\bibinfo{year}{1970}).

\bibitem[{\citenamefont{van~der Hoef and Madden}(1999)}]{Hoef:1999-1520}
\bibinfo{author}{\bibfnamefont{M.~A.} \bibnamefont{van~der Hoef}}
  \bibnamefont{and} \bibinfo{author}{\bibfnamefont{P.~A.}
  \bibnamefont{Madden}}, \bibinfo{journal}{J.\ Chem.\ Phys.}
  \textbf{\bibinfo{volume}{111}}, \bibinfo{pages}{1520} (\bibinfo{year}{1999}).

\bibitem[{\citenamefont{Louis}(2002)}]{Louis:2002-9187}
\bibinfo{author}{\bibfnamefont{A.~A.} \bibnamefont{Louis}},
  \bibinfo{journal}{J.\ Phys.\ Cond.\ Matter} \textbf{\bibinfo{volume}{14}},
  \bibinfo{pages}{9187} (\bibinfo{year}{2002}).

\bibitem[{\citenamefont{Fischer et~al.}(1984)\citenamefont{Fischer, Lustig,
  {Breitenfelder-Manske}, and Lemming}}]{Fischer:1984-485}
\bibinfo{author}{\bibfnamefont{J.}~\bibnamefont{Fischer}},
  \bibinfo{author}{\bibfnamefont{R.}~\bibnamefont{Lustig}},
  \bibinfo{author}{\bibfnamefont{H.}~\bibnamefont{{Breitenfelder-Manske}}},
  \bibnamefont{and} \bibinfo{author}{\bibfnamefont{W.}~\bibnamefont{Lemming}},
  \bibinfo{journal}{Mol.\ Phys.} \textbf{\bibinfo{volume}{52}},
  \bibinfo{pages}{485} (\bibinfo{year}{1984}).

\bibitem[{\citenamefont{Bembenek and Rice}(1999)}]{Bembenek:1999-1085}
\bibinfo{author}{\bibfnamefont{S.~D.} \bibnamefont{Bembenek}} \bibnamefont{and}
  \bibinfo{author}{\bibfnamefont{B.~M.} \bibnamefont{Rice}},
  \bibinfo{journal}{Mol.\ Phys.} \textbf{\bibinfo{volume}{97}},
  \bibinfo{pages}{1085} (\bibinfo{year}{1999}).

\bibitem[{\citenamefont{Vrabec et~al.}(2001)\citenamefont{Vrabec, Stoll, and
  Hasse}}]{Vrabec:2001-12126}
\bibinfo{author}{\bibfnamefont{J.}~\bibnamefont{Vrabec}},
  \bibinfo{author}{\bibfnamefont{J.}~\bibnamefont{Stoll}}, \bibnamefont{and}
  \bibinfo{author}{\bibfnamefont{H.}~\bibnamefont{Hasse}},
  \bibinfo{journal}{J.\ Phys.\ Chem. B} \textbf{\bibinfo{volume}{105}},
  \bibinfo{pages}{12126} (\bibinfo{year}{2001}).

\bibitem[{\citenamefont{Fernandez
  et~al.}(2004{\natexlab{a}})\citenamefont{Fernandez, Vrabec, and
  Hasse}}]{Fernandez:2004-175}
\bibinfo{author}{\bibfnamefont{J.~A.} \bibnamefont{Fernandez}},
  \bibinfo{author}{\bibfnamefont{J.}~\bibnamefont{Vrabec}}, \bibnamefont{and}
  \bibinfo{author}{\bibfnamefont{H.}~\bibnamefont{Hasse}},
  \bibinfo{journal}{Int.\ J.\ Thermophys.} \textbf{\bibinfo{volume}{25}}
  (\bibinfo{year}{2004}{\natexlab{a}}).

\bibitem[{\citenamefont{Fernandez
  et~al.}(2004{\natexlab{b}})\citenamefont{Fernandez, Vrabec, and
  Hasse}}]{Fernandez:2004-157}
\bibinfo{author}{\bibfnamefont{J.~A.} \bibnamefont{Fernandez}},
  \bibinfo{author}{\bibfnamefont{J.}~\bibnamefont{Vrabec}}, \bibnamefont{and}
  \bibinfo{author}{\bibfnamefont{H.}~\bibnamefont{Hasse}},
  \bibinfo{journal}{Fluid Phase Equilibria} \textbf{\bibinfo{volume}{221}},
  \bibinfo{pages}{157} (\bibinfo{year}{2004}{\natexlab{b}}).

\bibitem[{\citenamefont{Naugle}(1965)}]{Naugle:1965-3725}
\bibinfo{author}{\bibfnamefont{D.~G.} \bibnamefont{Naugle}},
  \bibinfo{journal}{J.\ Chem.\ Phys.} \textbf{\bibinfo{volume}{42}},
  \bibinfo{pages}{3725} (\bibinfo{year}{1965}).

\bibitem[{\citenamefont{Naugle et~al.}(1966)\citenamefont{Naugle, Lunsford, and
  Singer}}]{Naugle:1966-4669}
\bibinfo{author}{\bibfnamefont{D.~G.} \bibnamefont{Naugle}},
  \bibinfo{author}{\bibfnamefont{J.~H.} \bibnamefont{Lunsford}},
  \bibnamefont{and} \bibinfo{author}{\bibfnamefont{J.~R.}
  \bibnamefont{Singer}}, \bibinfo{journal}{J.\ Chem.\ Phys.}
  \textbf{\bibinfo{volume}{45}}, \bibinfo{pages}{4669} (\bibinfo{year}{1966}).

\bibitem[{\citenamefont{Naugle}(1966)}]{Naugle:1966-741}
\bibinfo{author}{\bibfnamefont{D.~G.} \bibnamefont{Naugle}},
  \bibinfo{journal}{J.\ Chem.\ Phys.} \textbf{\bibinfo{volume}{44}},
  \bibinfo{pages}{741} (\bibinfo{year}{1966}).

\bibitem[{\citenamefont{Swyt et~al.}(1967)\citenamefont{Swyt, Havlice, and
  Carome}}]{Swyt:1967-1199}
\bibinfo{author}{\bibfnamefont{D.~S.} \bibnamefont{Swyt}},
  \bibinfo{author}{\bibfnamefont{J.~F.} \bibnamefont{Havlice}},
  \bibnamefont{and} \bibinfo{author}{\bibfnamefont{E.~F.}
  \bibnamefont{Carome}}, \bibinfo{journal}{J.\ Chem.\ Phys.}
  \textbf{\bibinfo{volume}{47}}, \bibinfo{pages}{1199} (\bibinfo{year}{1967}).

\bibitem[{\citenamefont{Madigosky}(1967)}]{Madigosky:1967-4441}
\bibinfo{author}{\bibfnamefont{W.~M.} \bibnamefont{Madigosky}},
  \bibinfo{journal}{J.\ Chem.\ Phys.} \textbf{\bibinfo{volume}{46}},
  \bibinfo{pages}{4441} (\bibinfo{year}{1967}).

\bibitem[{\citenamefont{Cowan and Ball}(1972)}]{Cowan:1972-1881}
\bibinfo{author}{\bibfnamefont{J.~A.} \bibnamefont{Cowan}} \bibnamefont{and}
  \bibinfo{author}{\bibfnamefont{R.~N.} \bibnamefont{Ball}},
  \bibinfo{journal}{Can.\ J.\ Phys.} \textbf{\bibinfo{volume}{50}},
  \bibinfo{pages}{1881} (\bibinfo{year}{1972}).

\bibitem[{\citenamefont{Baharudin et~al.}(1975)\citenamefont{Baharudin,
  Jackson, Schoen, and Rouch}}]{Baharudin:1975-409}
\bibinfo{author}{\bibfnamefont{B.~Y.} \bibnamefont{Baharudin}},
  \bibinfo{author}{\bibfnamefont{D.~A.} \bibnamefont{Jackson}},
  \bibinfo{author}{\bibfnamefont{P.~E.} \bibnamefont{Schoen}},
  \bibnamefont{and} \bibinfo{author}{\bibfnamefont{J.}~\bibnamefont{Rouch}},
  \bibinfo{journal}{Phys.\ Lett.\ A} \textbf{\bibinfo{volume}{51}},
  \bibinfo{pages}{409} (\bibinfo{year}{1975}).

\bibitem[{\citenamefont{Malbrunot et~al.}(1983)\citenamefont{Malbrunot, Boyer,
  Charles, and Abachi}}]{Malbrunot:1983-1523}
\bibinfo{author}{\bibfnamefont{P.}~\bibnamefont{Malbrunot}},
  \bibinfo{author}{\bibfnamefont{A.}~\bibnamefont{Boyer}},
  \bibinfo{author}{\bibfnamefont{E.}~\bibnamefont{Charles}}, \bibnamefont{and}
  \bibinfo{author}{\bibfnamefont{H.}~\bibnamefont{Abachi}},
  \bibinfo{journal}{Phys.\ Rev.\ A} \textbf{\bibinfo{volume}{27}},
  \bibinfo{pages}{1523} (\bibinfo{year}{1983}).

\bibitem[{\citenamefont{Meier et~al.}(2005)\citenamefont{Meier, Laesecke, and
  Kabelac}}]{Meier:2005-014513}
\bibinfo{author}{\bibfnamefont{K.}~\bibnamefont{Meier}},
  \bibinfo{author}{\bibfnamefont{A.}~\bibnamefont{Laesecke}}, \bibnamefont{and}
  \bibinfo{author}{\bibfnamefont{S.}~\bibnamefont{Kabelac}},
  \bibinfo{journal}{J.\ Chem.\ Phys.} \textbf{\bibinfo{volume}{122}},
  \bibinfo{pages}{014513} (\bibinfo{year}{2005}).

\bibitem[{\citenamefont{Levesque et~al.}(1988)\citenamefont{Levesque, Weis, and
  Vermesse}}]{Levesque:1988-918}
\bibinfo{author}{\bibfnamefont{D.}~\bibnamefont{Levesque}},
  \bibinfo{author}{\bibfnamefont{J.~J.} \bibnamefont{Weis}}, \bibnamefont{and}
  \bibinfo{author}{\bibfnamefont{J.}~\bibnamefont{Vermesse}},
  \bibinfo{journal}{Phys.\ Rev.\ A} \textbf{\bibinfo{volume}{37}},
  \bibinfo{pages}{918} (\bibinfo{year}{1988}).

\bibitem[{\citenamefont{Levesque and Weis}(1988)}]{Levesque:1988-3967}
\bibinfo{author}{\bibfnamefont{D.}~\bibnamefont{Levesque}} \bibnamefont{and}
  \bibinfo{author}{\bibfnamefont{J.~J.} \bibnamefont{Weis}},
  \bibinfo{journal}{Phys.\ Rev.\ A} \textbf{\bibinfo{volume}{37}},
  \bibinfo{pages}{3967} (\bibinfo{year}{1988}).

\bibitem[{\citenamefont{Heyes}(1984)}]{Heyes:1984-1363}
\bibinfo{author}{\bibfnamefont{D.~M.} \bibnamefont{Heyes}},
  \bibinfo{journal}{J.\ Chem.\ Soc., Faraday Trans.}
  \textbf{\bibinfo{volume}{80}}, \bibinfo{pages}{1363} (\bibinfo{year}{1984}).

\bibitem[{\citenamefont{Levesque and Verlet}(1987)}]{Levesque:1987-143}
\bibinfo{author}{\bibfnamefont{D.}~\bibnamefont{Levesque}} \bibnamefont{and}
  \bibinfo{author}{\bibfnamefont{L.}~\bibnamefont{Verlet}},
  \bibinfo{journal}{Mol.\ Phys.} \textbf{\bibinfo{volume}{61}},
  \bibinfo{pages}{143} (\bibinfo{year}{1987}).

\bibitem[{\citenamefont{Hoheisel et~al.}(1987)\citenamefont{Hoheisel,
  Vogelsang, and Schoen}}]{Hoheisel:1987-7195}
\bibinfo{author}{\bibfnamefont{C.}~\bibnamefont{Hoheisel}},
  \bibinfo{author}{\bibfnamefont{R.}~\bibnamefont{Vogelsang}},
  \bibnamefont{and} \bibinfo{author}{\bibfnamefont{M.}~\bibnamefont{Schoen}},
  \bibinfo{journal}{J.\ Chem.\ Phys.} \textbf{\bibinfo{volume}{87}},
  \bibinfo{pages}{7195} (\bibinfo{year}{1987}).

\bibitem[{\citenamefont{Allen and Tildesley}(1991)}]{Allen:CSL}
\bibinfo{author}{\bibfnamefont{M.~P.} \bibnamefont{Allen}} \bibnamefont{and}
  \bibinfo{author}{\bibfnamefont{D.~J.} \bibnamefont{Tildesley}},
  \emph{\bibinfo{title}{Computer Simulation of Liquids}}
  (\bibinfo{publisher}{Oxford University Press}, \bibinfo{year}{1991}).

\bibitem[{\citenamefont{Green}(1954)}]{Green:1954-398}
\bibinfo{author}{\bibfnamefont{M.~S.} \bibnamefont{Green}},
  \bibinfo{journal}{J.\ Chem.\ Phys.} \textbf{\bibinfo{volume}{22}},
  \bibinfo{pages}{398} (\bibinfo{year}{1954}).

\bibitem[{\citenamefont{Lee and Cummings}(1994)}]{Lee:1994-6206}
\bibinfo{author}{\bibfnamefont{S.~H.} \bibnamefont{Lee}} \bibnamefont{and}
  \bibinfo{author}{\bibfnamefont{P.~T.} \bibnamefont{Cummings}},
  \bibinfo{journal}{J.\ Chem.\ Phys.} \textbf{\bibinfo{volume}{101}},
  \bibinfo{pages}{6206} (\bibinfo{year}{1994}).

\bibitem[{\citenamefont{Frenkel and Smit}(2002)}]{Frenkel:UMS}
\bibinfo{author}{\bibfnamefont{D.}~\bibnamefont{Frenkel}} \bibnamefont{and}
  \bibinfo{author}{\bibfnamefont{B.}~\bibnamefont{Smit}},
  \emph{\bibinfo{title}{Understanding Molecular Simulation}}
  (\bibinfo{publisher}{Academic Press}, \bibinfo{year}{2002}).

\bibitem[{\citenamefont{Marcelli et~al.}(2001)\citenamefont{Marcelli, Todd, and
  Sadus}}]{Marcelli:2001-021204}
\bibinfo{author}{\bibfnamefont{G.}~\bibnamefont{Marcelli}},
  \bibinfo{author}{\bibfnamefont{B.~D.} \bibnamefont{Todd}}, \bibnamefont{and}
  \bibinfo{author}{\bibfnamefont{R.~J.} \bibnamefont{Sadus}},
  \bibinfo{journal}{Phys.\ Rev. E} \textbf{\bibinfo{volume}{63}},
  \bibinfo{pages}{021204} (\bibinfo{year}{2001}).

\bibitem[{\citenamefont{Lifshitz and Pitaevskii}(1981)}]{Lifshitz:PK}
\bibinfo{author}{\bibfnamefont{E.~M.} \bibnamefont{Lifshitz}} \bibnamefont{and}
  \bibinfo{author}{\bibfnamefont{L.~P.} \bibnamefont{Pitaevskii}},
  \emph{\bibinfo{title}{Physical Kinetics}}
  (\bibinfo{publisher}{Butterworth-Heinemann}, \bibinfo{year}{1981}).

\bibitem[{\citenamefont{Choh and Uhlenbeck}(1958)}]{Choh:KTPDG}
\bibinfo{author}{\bibfnamefont{S.~T.} \bibnamefont{Choh}} \bibnamefont{and}
  \bibinfo{author}{\bibfnamefont{G.~E.} \bibnamefont{Uhlenbeck}},
  \emph{\bibinfo{title}{The kinetic theory of phenomena in dense gases}},
  \bibinfo{howpublished}{Navy Theoretical Physics, Contract No.~Nonr 1224 (15)}
  (\bibinfo{year}{1958}).

\bibitem[{\citenamefont{Baimbetov and Shaltykov}(1984)}]{Baimbetov:1984-1781}
\bibinfo{author}{\bibfnamefont{F.~B.} \bibnamefont{Baimbetov}}
  \bibnamefont{and} \bibinfo{author}{\bibfnamefont{N.~B.}
  \bibnamefont{Shaltykov}}, \bibinfo{journal}{Int.\ J.\ Heat Mass Transfer}
  \textbf{\bibinfo{volume}{27}}, \bibinfo{pages}{1781} (\bibinfo{year}{1984}).

\bibitem[{\citenamefont{Linstrom and Mallard}(retrieved May 31,
  2011)}]{NIST-old}
\bibinfo{editor}{\bibfnamefont{P.~J.} \bibnamefont{Linstrom}} \bibnamefont{and}
  \bibinfo{editor}{\bibfnamefont{W.~G.} \bibnamefont{Mallard}}, eds.,
  \emph{\bibinfo{title}{NIST Chemistry WebBook, NIST Standard Reference
  Database Number 69}} (\bibinfo{publisher}{National Institute of Standards and
  Technology}, \bibinfo{address}{Gaithersburg MD, 20899},
  \bibinfo{year}{retrieved May 31, 2011}),
  \urlprefix\url{http://webbook.nist.gov}.

\end{thebibliography}


\end{document}